\documentclass[11pt,a4paper]{article}
\usepackage{jcappub}
\usepackage{float}
\usepackage{rotating}
\usepackage{graphicx}
\usepackage{epsfig}
\usepackage{amsmath}
\usepackage{amssymb}
\usepackage{mathrsfs}  
\usepackage{subfigure}
\usepackage{time}
\usepackage{relsize, comment}
\usepackage{xcolor}
\usepackage[utf8]{inputenc}

\hypersetup{ linktoc=all,
    colorlinks=true, linkcolor={blue},  
       citecolor={red}, urlcolor={vividviolet}
}
\definecolor{rosy}{RGB}{230,235,252}
\definecolor{myframetitle}{RGB}{90,89,170}
\definecolor{myblocktitle}{RGB}{140,185,249}
\definecolor{mytitle}{RGB}{10,80,26}

\definecolor{darkgreen}{RGB}{27,130,45}
\definecolor{darkblue}{rgb}{0,0,0.3}
\definecolor{darkred}{rgb}{0.7,0,0}

\definecolor{light gray}{RGB}{220,220,220}
\definecolor{dark purple}{RGB}{108,0,217}
\definecolor{pink}{RGB}{190,20,100}
\definecolor{orang}{RGB}{193,63,0}
\definecolor{green}{RGB}{11,98,17}
\definecolor{darkpink}{RGB}{153,0,76}
\definecolor{bluegreen}{RGB}{0,102,102}
\definecolor{greenlagan}{RGB}{0,102,0}
\definecolor{redgreen}{RGB}{102,102,0}
\definecolor{Redgreen}{RGB}{153,76,0}
\definecolor{vividviolet}{rgb}{0.62, 0.0, 1.0}
\definecolor{amaranth}{rgb}{0.9, 0.17, 0.31}
\definecolor{palatinateblue}{rgb}{0.15, 0.23, 0.89}
\definecolor{brightpink}{rgb}{1.0, 0.0, 0.5}
\definecolor{cornflowerblue}{rgb}{0.39, 0.58, 0.93}
\definecolor{deepcarminepink}{rgb}{0.94, 0.19, 0.22}
\definecolor{radicalred}{rgb}{1.0, 0.21, 0.37}

\title{Gravitational wave probes of Barrow cosmology with LISA standard sirens}
\author[a,b]{Mahnaz Asghari}
\affiliation[a]{Department of Physics, College of Sciences, Shiraz University, Shiraz 71454, Iran}
\affiliation[b]{Biruni Observatory, Collage of Sciences, Shiraz University, Shiraz 71454, Iran}
\emailAdd{mahnaz.asghari@shirazu.ac.ir}
\author[c]{Alireza Allahyari}
\emailAdd{alireza.al@khu.ac.ir}
\affiliation[c]{Department of Astronomy and High Energy Physics, Kharazmi University, 15719-14911, Tehran, Iran \looseness=-1}
\author[d]{and David F. Mota}
\affiliation[d]{Institute of Theoretical Astrophysics, University of Oslo, P.O. Box 1029 Blindern, N-0315 Oslo, Norway}
\emailAdd{d.f.mota@astro.uio.no}

\abstract{We study the Barrow cosmological model, which proposes that quantum gravity effects create a complex, fractal structure for the universe's apparent horizon. We leverage the thermodynamics-gravity conjecture. By applying the Clausius relation to the apparent horizon of the Friedmann-Lemaître-Robertson-Walker universe within this framework, we derive modified field equations where the Barrow entropy is linked to the horizon. We assess the Barrow cosmology against current observations – cosmic microwave background , supernovae , and baryon acoustic oscillations data – and include projections for future Laser Interferometer Space Antenna (LISA) standard sirens (SS). Our numerical results suggest a modest improvement in the Hubble tension for Barrow cosmology with phantom dark energy behavior, compared to the standard cosmological model. Furthermore, incorporating simulated LISA SS data alongside existing observational constraints tightens the limitations on cosmological parameters, particularly the deformation exponent.}

\begin{document}
\maketitle
\flushbottom

	\section{Introduction}
	It is known that there is a remarkable connection between thermodynamics and gravity \cite{bh1,bh2,bh3}, where accordingly, the Einstein field equations can be considered as the equation of state for the spacetime \cite{j1}. From a cosmological perspective, the thermodynamics-gravity conjecture yields the Friedmann equations in the form of the first law of thermodynamics and vice versa \cite{fr1,fr2,fr3,fr4,fr5,fr6,fr7,fr8}. So, contemplating the spacetime as a thermodynamic system, we may write the field equations of Friedmann-Lema\^itre-Robertson-Walker (FLRW) universe by making use of the Clausius relation $\delta Q=T \delta S$ at the apparent horizon of universe. Correspondingly, applying Bekenstein-Hawking area law entropy \cite{bh4,bh3} in Clausius relation yields field equations in concordance $\Lambda$CDM model, while modified theories of gravity are reproduced by making corrections to the standard area law \cite{grth}. 
	
	Regarding modifications on area law entropy relation, there are logarithmic as well as power-law corrections to Bekenstein-Hawking entropy due to quantum effects, which have been extensively explored in the literature \cite{qc1,qc2,qc3,qc4,qc5,qc6,qc7,qc8,qc9,qc10,qc11,qc12}. Also, one can consider non-extensive Tsallis entropy \cite{ts1} proposed by the fact that Boltzmann-Gibbs additive entropy should be generalized to non-additive one in large scale gravitational systems \cite{bg1,na1,na2,na3,na4,na5,na6}. Related investigations on Tsallis entropy are developed in \cite{ts2,ts3,ts4,ts5,ts6,ts7,ts8,ts9,ts10,ts11,ts12,ts13,ts14,ts15,ts16,ts17,ts18,ts19,ts20,ts21,ts22,ts23,ts24,ts25}. 
	
	Furthermore, J. D. Barrow in 2020, introduced a fractal structure for the black hole horizon geometry due to quantum-gravitational effects \cite{bw1}. The Barrow formulation of black hole entropy is given by \cite{bw1}
	\begin{equation} \label{eq1}
	S=\Big(\frac{A}{A_0}\Big)^{1+\Delta/2} \;,
	\end{equation} 
	where $A$ stands for the black hole horizon area, $A_0$ indicates the Planck area, and the deformation exponent $\Delta$ ($0 \le \Delta \le 1$ \cite{bw1,bw2,bw3})  specifies deviations from the area law formula. Choosing $\Delta=0$ corresponds to the standard Bekenstein-Hawking entropy, and consequently shows the simplest horizon construction. On the other hand, $\Delta=1$ describes the most intricate and fractal structure. For some theoretical explorations on Barrow entropy in the cosmological context, see \cite{bw4,bw5,bw6,bw7,bw8,bw9,bw10,bw11,bw12,bw13,bw14,bw15,bw16,bw17,bw18,bw19,bw20,bw21,bw22,bw23,bw24,bw25}. Moreover, it is important to investigate Barrow cosmology from an observational viewpoint \cite{bwobs1,bwobs2,bwobs3,bwobs4,bwobs5,bwobs6,bwobs7}. In this direction, one powerful probe to inspect alternative cosmological theories to standard $\Lambda$CDM model is the gravitational wave (GW) astronomy.
	
	The direct detection of GW signal from the binary black hole coalescence GW150914 by the LIGO\footnote{Laser Interferometer Gravitational-Wave Observatory} collaboration in 2015 \cite{ligo}, provided significant advancements in investigating the gravitational physics. Thereafter in 2017, the LIGO-Virgo detector network identified a GW signal emitted by the coalescing binary neutron star GW170817 \cite{gw1}, which was associated with a short $\gamma$-ray burst GRB170817A \cite{gw2}, observed by the Fermi Gamma-Ray Burst Monitor \cite{fermi1,fermi2,fermi3,fermi4} and INTEGRAL \cite{integral1,integral2}. The intriguing coincident detection of the GW event and its electromagnetic (EM) counterpart inaugurated the era of multi-messenger astrophysics \cite{multi1}.  
	
	Considering cosmological applications, one can regard coalescing binary systems as standard sirens (SS) (i.e. the gravitational analogue of standard candles) which are utilized to estimate the Hubble constant \cite{schutz1,schutz2}. Actually it is possible to determine the luminosity distance to the compact binary coalescence by measuring its generated GW signal, and then the SS measurement of $d_L$ results in estimating cosmological parameters \cite{ss1,ss2,ss3}. In 2017, the first SS determination of $H_0$, which was independent of the cosmic distance ladder, reported the constraint of $H_0=70.0^{+12.0}_{-8.0}$ $\mathrm{km\,s^{-1}\,Mpc^{-1}}$ \cite{ssobs1}. Afterwards, another measurement of $H_0$ based on multiple GW observations resulted in $H_0=68.7^{+17.0}_{-7.8}$ $\mathrm{km\,s^{-1}\,Mpc^{-1}}$, in which according to a statistical method, informations from GW170817 along with its EM counterpart was combined with GW signals from binary black hole events with no expected EM counterparts \cite{ssobs2}. Regarding unsatisfactory accuracy on Hubble constant from SS measurements, it is important to exploit next generation GW observatories such as the ground-based detectors Einstein Telescope \cite{et1,et2} and Cosmic Explorer \cite{ce}, as well as the space-based detector LISA\footnote{Laser Interferometer Space Antenna} \cite{lisa1,lisa2} to enhance constraints on cosmological parameters.  
	
	The space-borne interferometer LISA, expected to be launched in 2034, has the potentiality to detect GWs in a band from below $10^{-4}$ Hz to above $10^{-1}$ Hz. The low frequency range of the LISA mission provides the opportunity to probe SS signals at much higher redshifts as well as exploring more massive black hole mergers \cite{lisa1}. There are several studies on applying SS in investigating the observational verification of cosmological models. For some recent related works see e.g. \cite{ssapp1,ssapp2,ssapp3,ssapp4,ssapp5,ssapp6,ssapp7,ssapp8,ssapp9,ssapp10,ssapp11,ssapp12,ssapp13,ssapp14,ssapp15,ssapp16,ssapp17,ssapp18,ssapp19,ssapp20,ssapp21}. 
	
	The present study is devoted to exploiting the LISA SS to put constraints on parameters of Barrow cosmological model. In particular, we apply GW SS together with cosmic microwave background (CMB), supernovae (SN), and baryon acoustic oscillations (BAO) measurements to examine the effectiveness of Barrow cosmology in resolving the Hubble tension \cite{H01,H02}. Principally, direct determinations of Hubble constant are in conflict with Planck CMB data \cite{H03,H04,H05,H06,Kamionkowski:2022pkx,Vagnozzi:2021tjv,Vagnozzi:2023nrq,
			Cervantes-Cota:2023wet,Riess:2024ohe}. For instance the most recent low redshift estimation of $H_0$ based on SH0ES\footnote{Supernovae and $\mathrm{H}_0$ for the Equation of State of dark energy} collaboration results in $H_0=73.30\pm1.04\,\mathrm{km\,s^{-1}\,Mpc^{-1}}$ \cite{H07}, which reports a $5\sigma$ difference with Planck measurements \cite{p18}. Then, considering cosmological tensions, it is so pertinent to explore new physics beyond the concordance $\Lambda$CDM model. 
	The methodology of our investigation is similar to the paper \cite{bwobs4}, in which we detected a slight reduction in $H_0$ tension.     
	
	This paper is organized as follows. Section \ref{sec2} considers modified field equations based on Barrow corrections to the area law entropy, as well as introducing modifications to GW propagation in Barrow model. Section \ref{sec3} is dedicated to observational probes employed in our study, which consists of current observations along with future LISA SS data. We also present the applied method for generating the GW mock catalog in section \ref{sec3}. Numerical results as well as observational constraints on Barrow model are reported in Section \ref{sec4}. We conclude  in section \ref{sec5}. Throughout the paper we consider the units where $k_\mathrm{B}=c=\hbar=1$. 
	\section{Barrow cosmology} \label{sec2}
	This section is dedicated to describing modified field equations according to Barrow entropy corrections, as well as introducing modifications to GW propagation in Barrow cosmology.
	\subsection{Modified field equations} 
	In pursuance of extracting field equations from Barrow entropy, one should apply the Clausius relation $\delta Q=T \delta S$ at the apparent horizon of FLRW universe, which is considered as a whole thermodynamic system. 
	For this purpose, we regard a spatially flat universe, described by the FLRW metric	 
	\begin{equation} \label{eq2}
	\mathrm{d}s^2=a^2(\tau)\Big(-\mathrm{d}\tau^2+\mathrm{d}\vec{x}^2\Big) \;,
	\end{equation} 
	in background level. 
	Then, it is also important to contemplate the area $A$ in Barrow entropy relation (\ref{eq1}) as the apparent horizon of the universe area $A=4\pi \tilde{r}^2_\mathrm{A}$, with the apparent horizon radius $\tilde{r}_\mathrm{A}=\Big(H^2+\frac{K}{a^2}\Big)^{-1/2}$ \cite{ra1} (where $H$ is the Hubble parameter, and $K=-1,0,1$ is the curvature constant describing open, flat and closed universe, respectively). 
	 
	Taking the differential of Barrow entropy relation (\ref{eq1}), results in 
	\begin{equation} \label{eq3}
	\delta S=\Big(1+\frac{\Delta}{2}\Big) A_0^{-1-\Delta/2} A^{\Delta/2} \delta A \;.
	\end{equation}
	And additionally, following the Refs. \cite{j1,grth}, $\delta Q$ is defined as
	\begin{equation} \label{eq4}
	\delta Q= -\kappa \int_{\mathcal{H}}^{} \lambda T_{\mu \nu} k^{\mu} k^{\nu} \mathrm{d}\lambda \mathrm{d}A \;,
	\end{equation}
	in which $\mathcal{H}$ stands for the apparent horizon, $\delta A=\int_{\mathcal{H}}^{} \theta \mathrm{d}\lambda \mathrm{d}A$, and $\theta=-\lambda R_{\mu \nu} k^{\mu} k^{\nu}$. Also the content of universe including radiation (R), matter (M) (dark matter (DM) and baryons (B)) and dark energy (DE), is considered as a perfect fluid with the energy-momentum tensor $T_{\mu \nu}=\big(\rho+p\big)u_{\mu}u_{\nu}+g_{\mu \nu}p$. Furthermore, the temperature corresponds to the apparent horizon with the surface gravity $\kappa$ is 
	\begin{equation} \label{eq5}	
	T=\frac{\kappa}{2\pi} \;.
    \end{equation}
	Replacing equations (\ref{eq3}), (\ref{eq4}) and (\ref{eq5}) in Clausius relation, yields
	\begin{equation} \label{eq6}
	\int_{\mathcal{H}}^{} (-\lambda) \Bigg(-2\pi T_{\mu \nu}+\Big(1+\frac{\Delta}{2}\Big) A_0^{-1-\Delta/2} R_{\mu \nu} A^{\Delta/2} \Bigg) k^{\mu} k^{\nu} \mathrm{d}\lambda \mathrm{d}A=0 \;, 
	\end{equation}
	where by introducing the scalar $f$, for all null vectors $k^{\mu}$ we obtain
	\begin{equation} \label{eq7}
	-2\pi T_{\mu \nu}+\Big(1+\frac{\Delta}{2}\Big) A_0^{-1-\Delta/2} R_{\mu \nu} A^{\Delta/2}=f g_{\mu \nu} \;.
	\end{equation}
	Then, imposing the energy-momentum conservation results in	
	\begin{equation} \label{eq8}
	\Big(1+\frac{\Delta}{2}\Big) A_0^{-1-\Delta/2} \Bigg(\frac{1}{2}\big(\partial_{\nu} R\big) A^{\Delta/2} + R_{\mu \nu} \partial^{\mu} A^{\Delta/2}\Bigg) =\partial_{\nu} f \,.
	\end{equation}
	So, since the LHS of (\ref{eq8}) is not the gradient of a scalar, one can conclude that on the account of non-equilibrium thermodynamics, the Clausius relation is not satisfied. Accordingly, we apply the entropy balance relation \cite{grth}
	\begin{equation} \label{eq9}
	\delta S=\frac{\delta Q}{T}+\mathrm{d}_i S \;,
	\end{equation}
	in which $\mathrm{d}_i S$ is the produced entropy owing to irreversible processes inside the system \cite{en1}. Then, contemplating the energy-momentum conservation, $\mathrm{d}_i S$ can be written as 
	\begin{equation} \label{eq10}
	\mathrm{d}_i S=\Big(1+\frac{\Delta}{2}\Big) A_0^{-1-\Delta/2} \int_{\mathcal{H}}^{} (-\lambda) \nabla_{\mu} \nabla_{\nu} A^{\Delta/2} k^{\mu} k^{\nu} \mathrm{d}\lambda \mathrm{d}A \;.
	\end{equation}
	Thereupon, by substituting $\mathrm{d}_i S$ in the entropy balance relation (\ref{eq9}) and repeating the above calculations (refer to the Ref. \cite{bwobs4} for more details), finally we find 
	\begin{equation}  \label{eq11}
	\Big(1+\frac{\Delta}{2}\Big) A_0^{-1-\Delta/2} \Bigg(\frac{1}{2} \big(\partial_{\nu} R\big) A^{\Delta/2}-\partial_{\nu} \Box A^{\Delta/2}\Bigg)=\partial_{\nu} f \;.
	\end{equation}
	Now, considering the scalar $\mathcal{L}$ as $\mathcal{L}=R A^{\Delta/2}$, yields
	\begin{align} \label{eq12}
	& \frac{\partial \mathcal{L}}{\partial R}=A^{\Delta/2} \;, \nonumber\\
	& \partial_{\nu} R=\frac{\partial R}{\partial x^{\nu}}=\frac{\partial R}{\partial \mathcal{L}}\frac{\partial \mathcal{L}}{\partial x^{\nu}}=A^{-\Delta/2} \partial_{\nu}\mathcal{L} \;,  \nonumber\\
	& \to \big(\partial_{\nu} R\big) A^{\Delta/2}=\partial_{\nu}\mathcal{L} \;,
	\end{align}
	and accordingly we find the scalar $f$ as
	\begin{equation}  \label{eq13}
	f=\Big(1+\frac{\Delta}{2}\Big) A_0^{-1-\Delta/2} \Big(\frac{1}{2} R A^{\Delta/2} - \Box A^{\Delta/2}\Big) \;.
	\end{equation}
	Thereupon, modified field equations based on Barrow entropy take the form
	\begin{equation} \label{eq14}
	R_{\mu \nu} A^{\Delta/2}-\nabla_{\mu} \nabla_{\nu} A^{\Delta/2}-\frac{1}{2} R A^{\Delta/2} g_{\mu \nu}+\Box A^{\Delta/2} g_{\mu \nu}=\frac{4\pi}{2+\Delta}A_0^{1+\Delta/2}T_{\mu \nu} \;.
	\end{equation}
	Consequently, by applying the Clausius relation at the apparent horizon of FLRW universe along with regarding quantum-gravitational effects on the apparent horizon, modified gravitational field equations in Barrow cosmology have been extracted. It can be easily seen that choosing $\Delta=0$ reduces to the standard field equations in general relativity.
	Contemplating $A^{\Delta/2}=(4\pi)^{\Delta/2} H^{-\Delta}$ for a spatially flat universe, modified field equations in background level can be derived as
	\begin{equation} \label{eq15}
		H^{2-\Delta}-\Delta H^{-\Delta} H' \frac{1}{a}=X^\Delta\,\frac{8\pi G}{3} \sum_{i}\bar{\rho}_i \;,
	\end{equation}
	\begin{equation} \label{eq16}
		H^{-\Delta}\Bigg{\{}\big(\Delta-2\big) H' \frac{1}{a}-3H^2+\Delta\bigg(\frac{H''}{H}-\big(1+\Delta\big)\Big(\frac{H'}{H}\Big)^2\bigg)\frac{1}{a^2}\Bigg{\}}=X^\Delta\,8\pi G \sum_{i}\bar{p}_i \;,
	\end{equation} 
	where a prime indicates a deviation with respect to the conformal time, and also $A_0$ is considered as 
	\begin{equation*}
		A_0=\big(4\pi\big)^{\Delta/(2+\Delta)}\Big(2\big(2+\Delta\big)G\Big)^{2/(2+\Delta)}X^{2\Delta/(2+\Delta)} \;,
	\end{equation*}
	with the constant $X$ of the dimension of length.
	Considering equations (\ref{eq15}) and (\ref{eq16}), the modified Friedmann equation in regard to Barrow entropy corrections takes the form
	\begin{equation} \label{eq17}
	H^{2-\Delta}=\frac{1}{1+2\Delta}X^\Delta\,\frac{8\pi G}{3}\sum_{i}\bar{\rho}_i \;.
	\end{equation}
	Furthermore, modified Friedmann equation in term of total density parameter $\Omega_\mathrm{tot}=\bar{\rho}_\mathrm{tot}/\rho_\mathrm{cr}$ (with $\rho_\mathrm{cr}={3H^2}/{(8\pi G)}$ and $\bar{\rho}_\mathrm{tot}=\sum_{i}\bar{\rho}_i$) becomes
	\begin{equation} \label{eq18}
		\Omega_\mathrm{tot}=\big(1+2\Delta\big)X^{-\Delta}H^{-\Delta} \;.
	\end{equation}
	Then, as expected, Friedmann equation in standard cosmology will be recovered in case of $\Delta=0$.
	
	It is also interesting to notice that the cosmic accelerated expansion in Barrow cosmology is achieved when \mbox{$w_\mathrm{tot}<-(1+\Delta)/3$}. So, contemplating the allowed range of $\Delta$ ($0 \le \Delta \le 1$), the most intricate structure of apparent horizon imposes the condition $w_\mathrm{tot}<-2/3$, compared to the standard gravity in which we have $w_\mathrm{tot}<-1/3$. It implies that Barrow cosmology late time acceleration will occur for more negative values of the equation of state parameter.
	
	Considering cosmological perturbations in FLRW universe, for the scalar mode of metric perturbations we have
	\begin{equation} 
	\mathrm{d}s^2=a^2(\tau)\Big(-\mathrm{d}\tau^2+\big(\delta_{ij}+h_{ij}\big)\mathrm{d}x^i\mathrm{d}x^j\Big) \;,
	\end{equation} 
	where $h_{ij}(\vec{x},\tau)=\int \mathrm{d}^3k\,e^{i\vec{k}.\vec{x}}
	\bigg(\hat{k}_i\hat{k}_jh(\vec{k},\tau)+\Big(\hat{k}_i\hat{k}_j-\frac{1}{3}\delta_{ij}\Big)6\eta(\vec{k},\tau)\bigg)$, with scalar perturbations $h$ and $\eta$, and $\vec{k}=k\hat{k}$ \cite{pt}.	
	So modified field equations (\ref{eq14}) to linear order of perturbations take the form
	\begin{equation} \label{eq19}
	\Big(\frac{a'}{a^2}\Big)^{-\Delta}\Bigg{\{}\frac{a'}{a}h'-2k^2\eta-\frac{1}{2}\Delta h'\bigg(\frac{a''}{a'}-2\frac{a'}{a}\bigg)\Bigg{\}}=X^\Delta\,a^2\,8\pi G \sum_{i}\delta \rho_{i} \;,
	\end{equation}
	\begin{equation} \label{eq20}
	\Big(\frac{a'}{a^2}\Big)^{-\Delta}\, k^2\eta'=X^\Delta\,a^2\,4\pi G \sum_{i}\big(\bar{\rho}_i+\bar{p}_i\big)\theta_{i} \;,
	\end{equation}
	\begin{align} \label{eq21}
	\Big(\frac{a'}{a^2}\Big)^{-\Delta}\Bigg{\{}\frac{1}{2}h''+3\eta''+\frac{a'}{a}h'+6\frac{a'}{a}\eta'-k^2\eta -\frac{1}{2}\Delta\bigg(\frac{a''}{a'}-2\frac{a'}{a}\bigg)\bigg(h'+6\eta'\bigg)\Bigg{\}}=0 \;,
	\end{align}
	\begin{equation} \label{eq22}
	\Big(\frac{a'}{a^2}\Big)^{-\Delta}\Bigg{\{}-2\frac{a'}{a}h'-h''+2k^2\eta+\Delta\bigg(\frac{a''}{a'}-2\frac{a'}{a}\bigg)h'\Bigg{\}}=X^\Delta\,a^2\,24\pi G \sum_{i}\delta p_{i} \;.
	\end{equation}	
	Moreover, the energy-momentum conservation equations are irrelevant to Barrow entropy corrections, so continuity and Euler equations for matter and dark energy components are given by
	\begin{align}
	\delta'_\mathrm{M}=-\theta_\mathrm{M}-\frac{1}{2}h' \;, \label{eq23}
	\end{align}
	\begin{align}
	\theta'_\mathrm{M}=-\frac{a'}{a}\theta_\mathrm{M} \;, \label{eq24}
	\end{align}
	\begin{align}
	\delta'_\mathrm{DE}&=-3\frac{a'}{a}\big(c^2_{s,\mathrm{DE}}-w_\mathrm{DE}\big)\delta_\mathrm{DE}-\frac{1}{2}h'\big(1+w_\mathrm{DE}\big) \nonumber \\
	&-\big(1+w_\mathrm{DE}\big)\bigg(1+9\Big(\frac{a'}{a}\Big)^2\big(c^2_{s,\mathrm{DE}}-c^2_{a,\mathrm{DE}}\big)\frac{1}{k^2}\bigg)\theta_\mathrm{DE} \;, \label{eq25}
	\end{align}
	\begin{align}
	\theta'_\mathrm{DE}=\frac{a'}{a}\big(3c^2_{s,\mathrm{DE}}-1\big)\theta_\mathrm{DE}+\frac{k^2c^2_{s,\mathrm{DE}}}{1+w_\mathrm{DE}}\delta_\mathrm{DE} \;. \label{eq26}
	\end{align}
	\subsection{Modified gravitational wave propagation}
	In this part, we turn our attention to the modified propagation of GWs in accordance with Barrow corrections to the area law entropy. To this end, we contemplate the tensorial part of perturbed field equations (\ref{eq14}) in absence of anisotropic stress, which reads
	\begin{equation} \label{eq27}
	h''_{(+,\times)}+\Big(2\frac{a'}{a}-a^2\Delta\big(\frac{a''}{a'}-2\frac{a'}{a}\big)\Big)h'_{(+,\times)}+k^2h_{(+,\times)}=0 \;,
	\end{equation}
	with the two plus and cross polarizations. From equation (\ref{eq27}) it can be understood that the friction term is modified, while the propagation speed of GWs remains unchanged which is compatible with observations \cite{gw2}. Also it is obvious that for $\Delta=0$, the free propagation of tensor perturbations in general relativity will be recovered. In order to solve equation (\ref{eq27}), it is convenient to consider the function $\chi_{(+,\times)}(\tau,k)$ defined as
	\begin{equation} \label{eq28}
	\chi_{(+,\times)}(\tau,k)=\tilde{a}(\tau) h_{(+,\times)}(\tau,k) \;,
	\end{equation}
	where in our modified cosmological model, $\tilde{a}(\tau)$ is given by
	\begin{equation} \label{eq29}
	\frac{\tilde{a}'}{\tilde{a}}=\frac{a'}{a}-a^2\Delta\bigg(\frac{1}{2}\frac{a''}{a'}-\frac{a'}{a}\bigg)=Ha-\frac{1}{2}a^2\Delta\frac{H'}{H} \;.
	\end{equation}
	Thus, it can be inferred that in modified theories of gravity, the GW amplitude decreases as $1/{\tilde{a}}$ during propagation through cosmological distances. Hence, it is suggested to consider the GW luminosity distance $d_L^\mathrm{gw}(z)$ in modified gravity which is defined as \cite{dlgw1,ssapp1}
	\begin{equation} \label{eq32}
	d_L^\mathrm{gw}(z)=\frac{1}{(1+z)\tilde{a}(z)}d_L^\mathrm{em}(z) \;,
	\end{equation}
	where $d_L^\mathrm{em}(z)=(1+z)\int_0^z{\frac{\mathrm{d}z}{H(z)}}$ is the standard luminosity distance. Then, from equation (\ref{eq29}) in Barrow cosmology we obtain
	\begin{equation} \label{eq33}
	d_L^\mathrm{gw}(z)=d_L^\mathrm{em}(z) \exp{\bigg(\frac{1}{2}\Delta \int_0^z{\mathrm{d}z\,\frac{1}{(1+z)^2H(z)}\frac{\mathrm{d}H(z)}{\mathrm{d}z}}\bigg)} \;,
	\end{equation}
	with the Hubble parameter $H(z)$ defined in equation (\ref{eq17}).
	
	In the next section, we introduce the observational tests employed in this investigation to put constraints on the parameters of Barrow cosmological model.
	\section{Observational probes} \label{sec3}
	In order to confront Barrow model with observations, we apply the following current and future data in our numerical analysis.
	\begin{itemize} 
	\item \textbf{CMB:} We contemplate CMB measurements form Planck 2018 data release, which consists of high-$l$ TT,TE,EE, low-$l$ EE, low-$l$ TT, and lensing data \cite{p18}.
	\item \textbf{SN:} Considering the type Ia supernovae (SN Ia) distance measurements, we take into account the Pantheon sample containing 1048 data points in the range of $0.01<z<2.3$ \cite{pan}.
	\item \textbf{BAO:} We make use of the 6dF Galaxy Survey (6dFGS) BAO detection which provides constraining the distance-redshift relation at $z_\mathrm{eff}=0.106$ \cite{bao1}, and also we consider the SDSS\footnote{Sloan Digital Sky Survey}-III BOSS\footnote{Baryon Oscillation Spectroscopic Survey} data which include LOWZ ($z_\mathrm{eff}=0.32$) and CMASS ($z_\mathrm{eff}=0.57$) galaxy samples \cite{bao2}.
	\item \textbf{LISA standard sirens:} The space-based detector LISA has the potential to probe signals from SS of massive black hole binaries (MBHBs) in the range $10^3$ to $10^7$ solar masses \cite{lisa1,mass}. MBHBs are powerful sources of GWs in the universe, as they are supposed to merge at the center of galaxies, which are gas-rich nuclear environments, and so these events are expected to produce a detectable EM counterpart. Moreover, detecting MBHBs in the center of galaxies is so useful to improve our understanding of galaxy evolution and cosmic structure formation. 
		
	In order to generate our catalog of SS events, we apply the methodology outlined in \cite{ssapp11}, which considers the redshift distribution of SS according to the explanations in Ref. \cite{redshd}. Based on competing scenarios for the initial conditions for the massive black hole formation and evolution at high redshifts, there are three distinct populations of MBHBs, namely Pop III, Delay, and No Delay population models \cite{catalog}. Pop III model is a light-seed scenario in which massive black holes are considered to form from the remnants of population III stars. On the other hand, the heavy-seed scenario assumes that massive black holes evolve from the collapse of protogalactic disks, leading to Delay (where a finite time delay between the merger of host galaxies and black hole coalescence is assumed), and No Delay (with considering no delays between the galaxy and black hole mergers) models \cite{catalog}. 
	
	It is also important to compute the realistic $1\sigma$ luminosity distance error to MBHBs, according to the LISA sensitivity, which can be estimated as \cite{error,ssapp11}
	\begin{equation}
     \sigma_\mathrm{LISA}^2=\sigma_\mathrm{delens}^2+\sigma_\mathrm{v}^2+\sigma_\mathrm{inst}^2+\bigg(\frac{\mathrm{d}}{\mathrm{d}z}(d_L)\sigma_\mathrm{photo}\bigg)^2 \;,
	\end{equation}    
	which consists of error contributions from weak lensing, peculiar velocity, LISA instrument, and redshift measurements. 	
	The weak lensing contribution is given by
	\begin{equation}
	\sigma_\mathrm{delens}(z)=F_\mathrm{delens}(z)\sigma_\mathrm{lens}(z) \;,
	\end{equation}
	where 
	\begin{align}
	& F_\mathrm{delens}(z)=1-\frac{0.3}{\pi/2}\arctan\bigg(\frac{z}{0.073}\bigg) \;, \\
	& \frac{\sigma_\mathrm{lens}(z)}{d_L(z)}=0.066\bigg(\frac{1-(1+z)^{-0.25}}{0.25}\bigg)^{1.8} \;.
	\end{align}
	The peculiar velocity uncertainty contribution takes the form
	\begin{equation}
	\frac{\sigma_\mathrm{v}(z)}{d_L(z)}=\bigg(1+\frac{c(1+z)^2}{H(z)d_L(z)}\bigg)\frac{500\,\mathrm{km/s}}{c} \;.
	\end{equation}
	The LISA instrumental error is 
	\begin{equation}
	\frac{\sigma_\mathrm{inst}(z)}{d_L(z)}=0.05\bigg(\frac{d_L(z)}{36.6\,\mathrm{Gpc}}\bigg) \;.
	\end{equation}
	And finally, for redshift measurement error we have 
	\begin{equation}
	\sigma_\mathrm{photo}(z)=0.03(1+z) \;, \quad \text{if} \;\; z>2 \;. 
	\end{equation}
	
	Throughout this study, we generate an SS mock catalog for each MBHB population, i.e. Pop III, Delay, and No Delay models, considering a ten-year LISA mission. The fiducial model is assumed to be the Barrow cosmology described in section \ref{sec2}, where for the cosmological parameters we have considered the obtained best fit values based on the "CMB+SN+BAO" dataset displayed in table \ref{tab1}. Accordingly, figure \ref{fig1} depicts the evolution of GW luminosity distance with redshift for the three LISA mock catalogs.
	\begin{figure}[ht!]
		\centering
		\includegraphics[width=8.5cm]{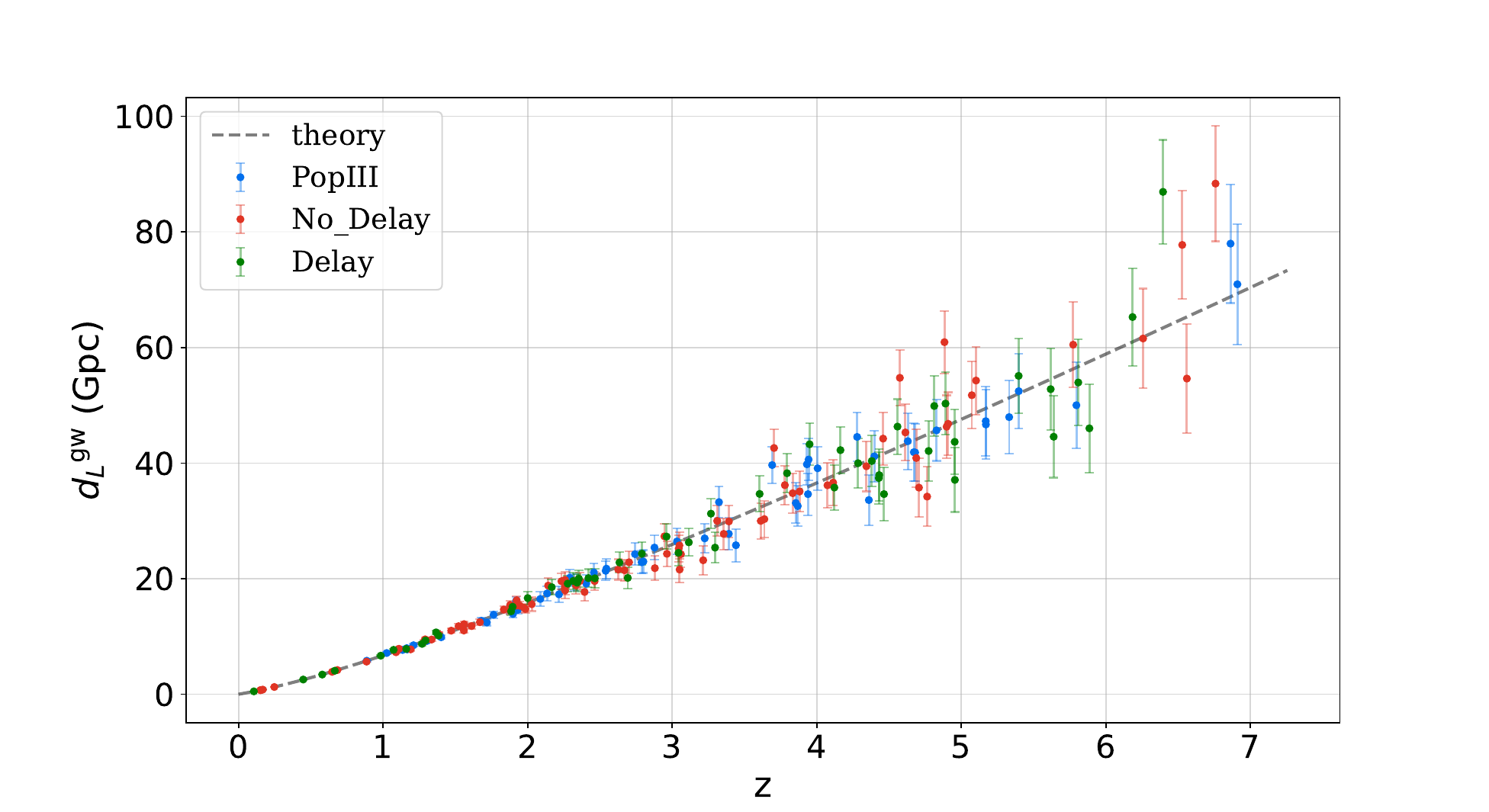} 
		\caption{Mock LISA SS for ten years of operation from three MBHB populations.}
		\label{fig1}
	\end{figure}
    \end{itemize}
    
    Next section is devoted to numerical results as well as observational constraints on Barrow cosmology based on the introduced data sets in this part.	
    \section{Results} \label{sec4}
    In the interest of a numerical investigation on Barrow cosmology, we make use of our modified version of the publicly available Boltzmann code CLASS\footnote{Cosmic Linear Anisotropy Solving System} \cite{cl1}, based on field equations of Barrow cosmological model. To this end, the value of cosmological parameters are set to the Planck 2018 data release \cite{p18}, while in case of dark energy component we consider $w_\mathrm{DE}=-0.98$ (to avoid some divergences in the dark energy perturbation equations) and $c_{s,\mathrm{DE}}^2=1$ (to simply have a smooth dark energy fluid).
    
\begin{figure}[ht!]
    	\includegraphics[width=8cm]{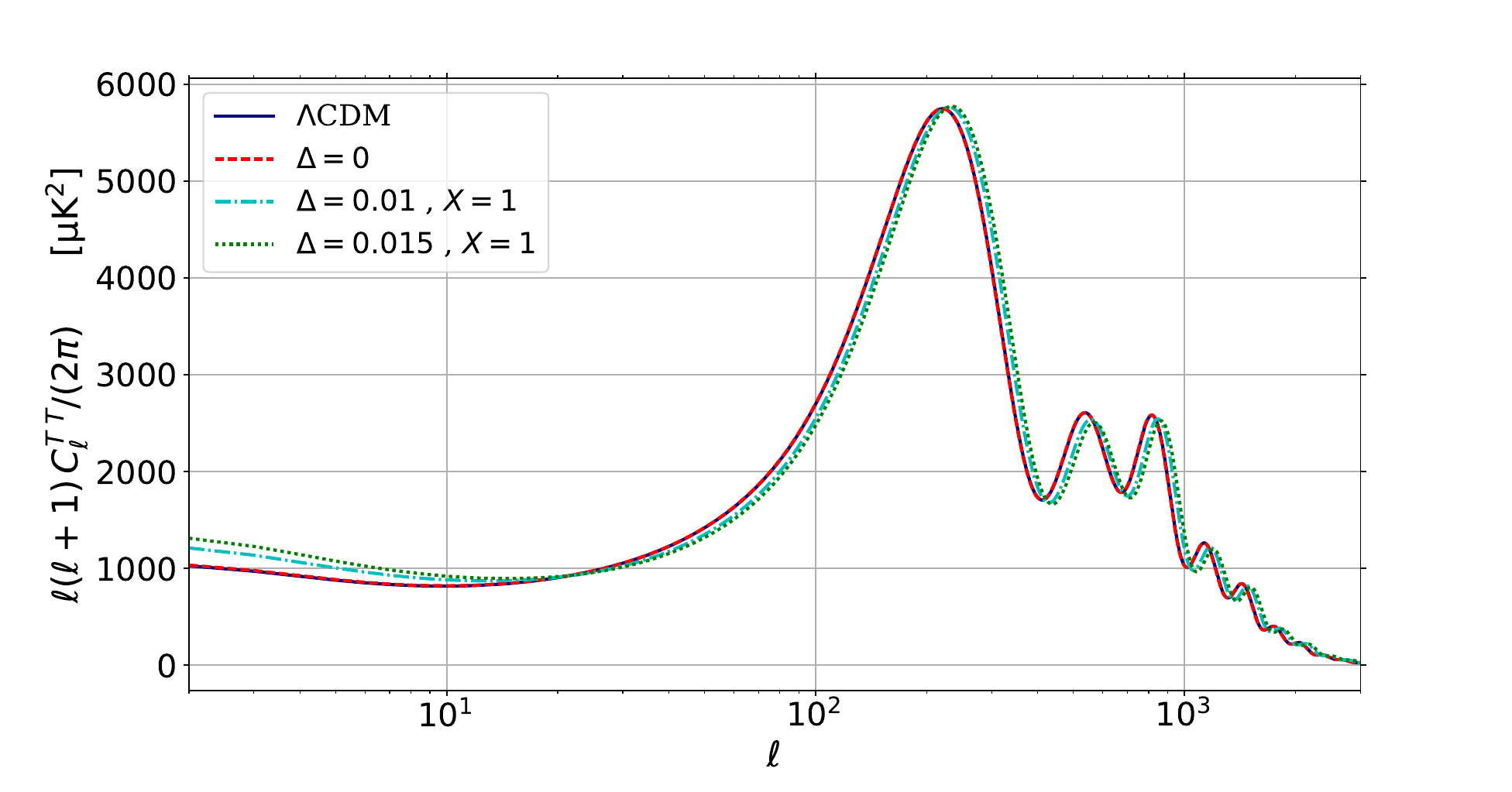} 
    	\includegraphics[width=8cm]{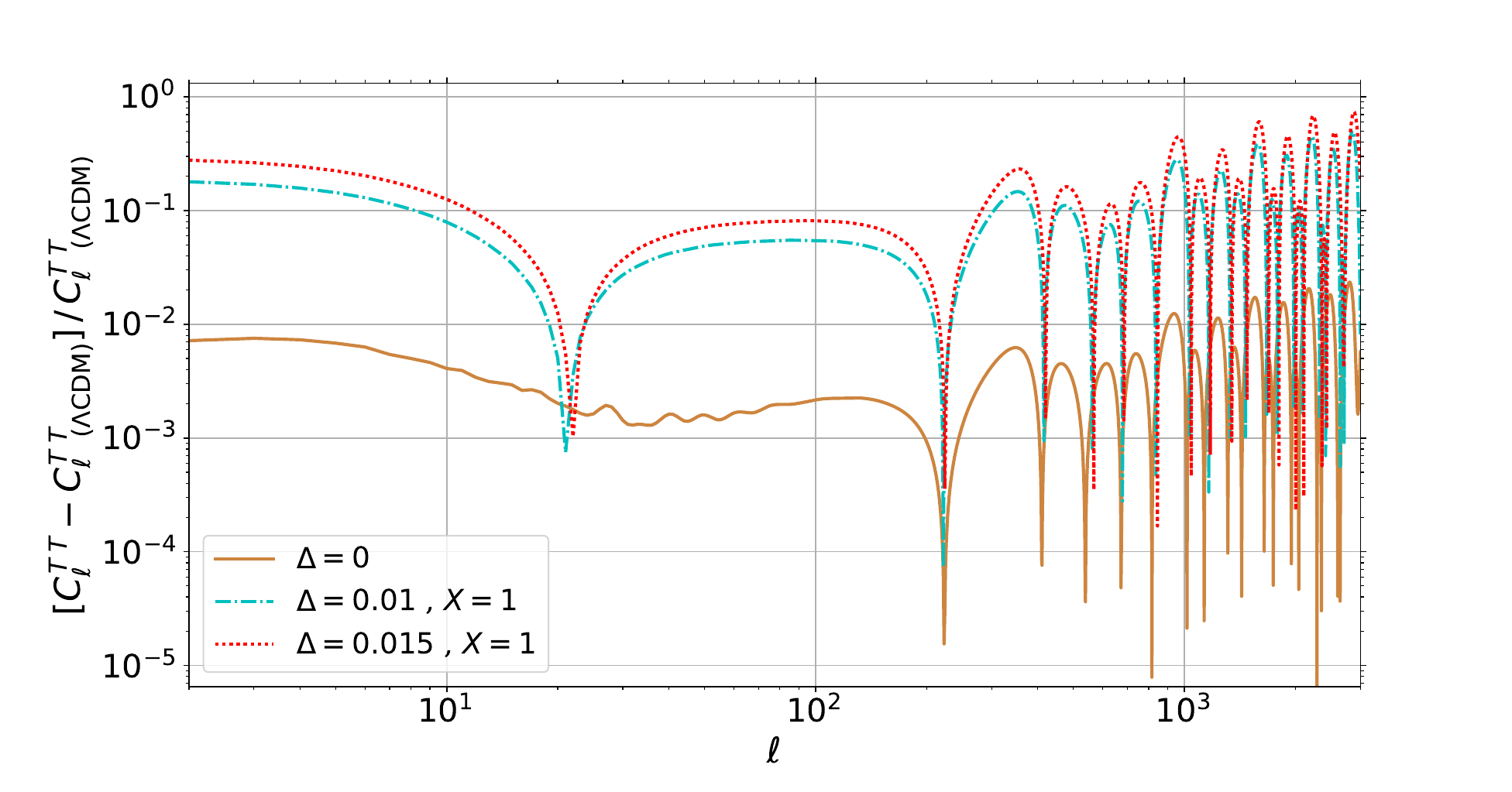} 
    	\includegraphics[width=8cm]{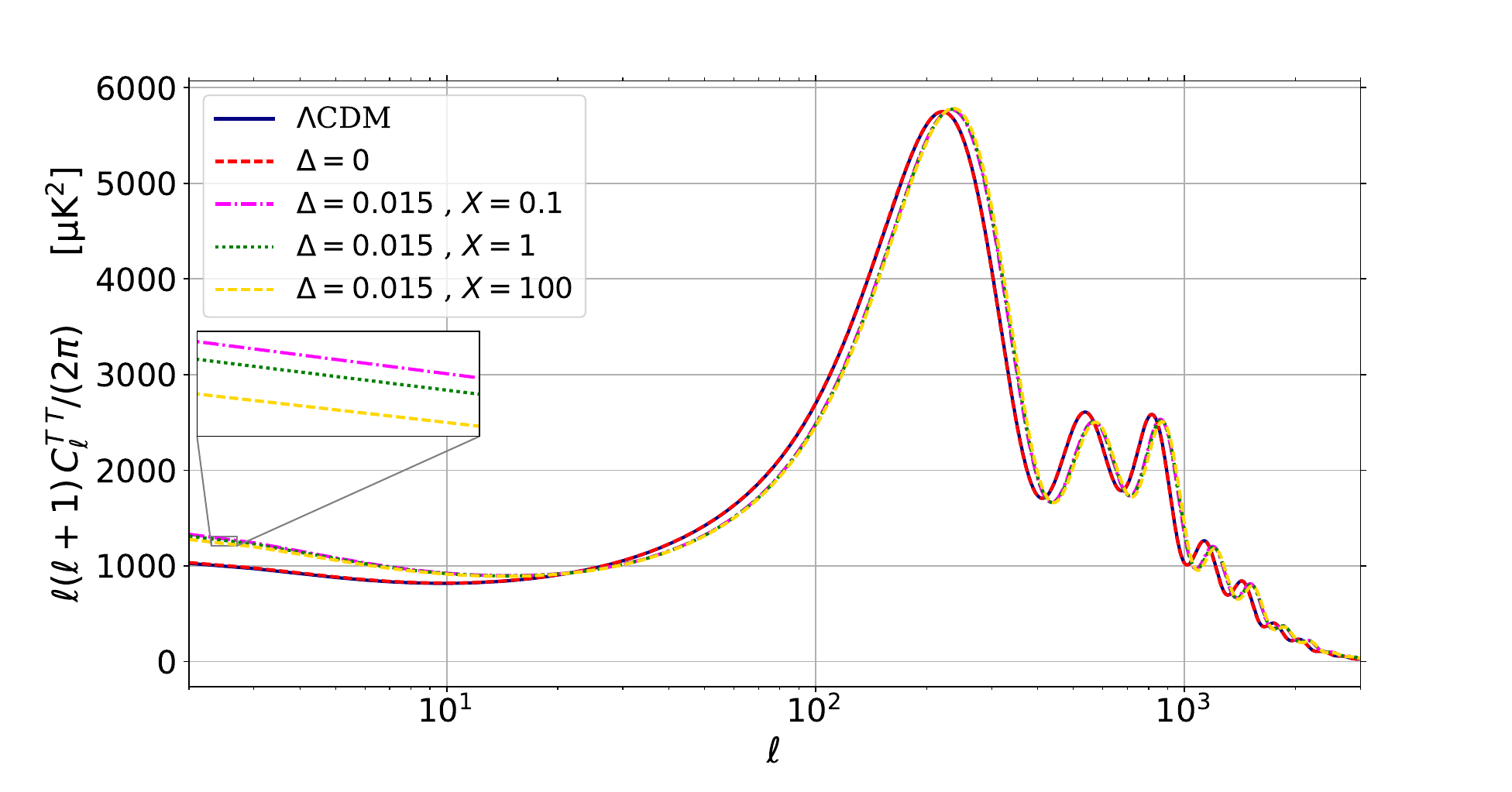} 
    	\includegraphics[width=8cm]{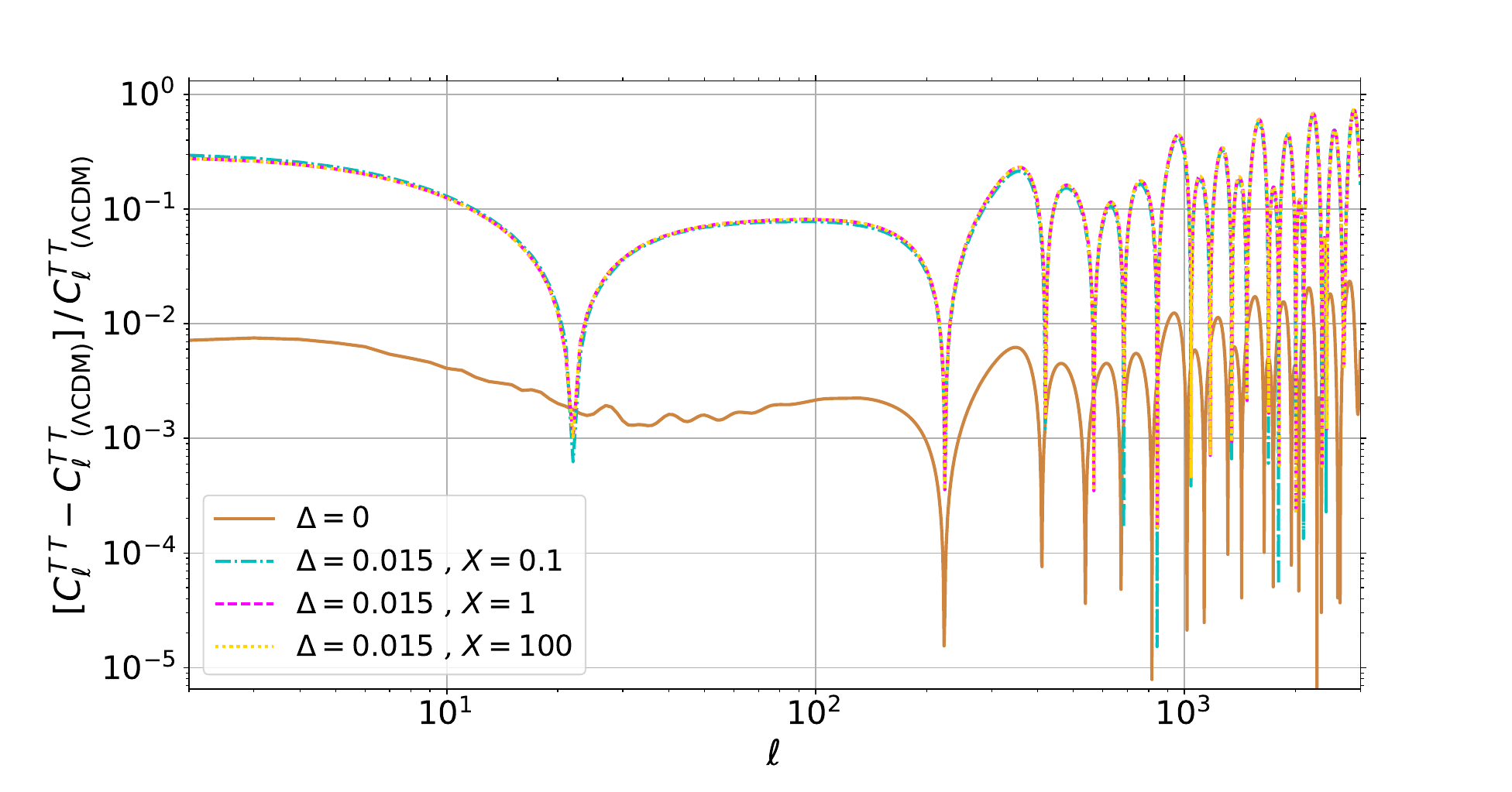}  
    	\caption{Upper panels: The CMB temperature power spectra (left) and their relative ratio with respect to standard cosmological model (right) for different values of $\Delta$, regarding $X=1$. Lower panels: The analogous diagrams for different values of $X$, where $\Delta=0.015$.}
    	\label{fig2}
    \end{figure}

    The CMB temperature power spectra diagrams in Barrow cosmology are illustrated in figure \ref{fig2}. Upper panels represent the power spectra for different values of $\Delta$, where $X$ fixed to be $1$, while lower panels depict the same diagrams for different values of $X$, with $\Delta$ considered to be $0.015$. (Notice that the case $X=1$ recovers the results reported in Ref. \cite{bwobs4}.)
     
\begin{figure}
		\includegraphics[width=8cm]{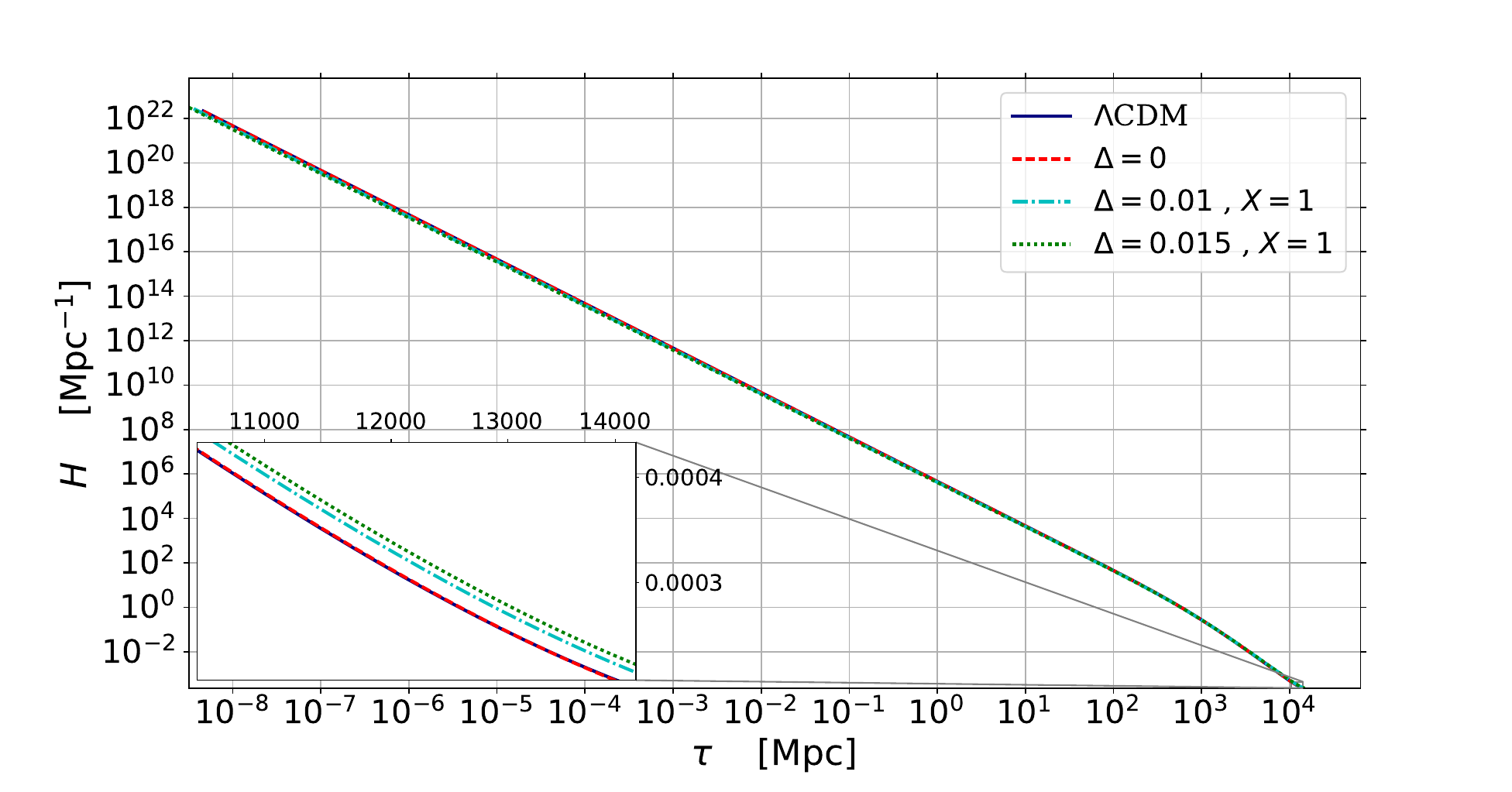}
		\includegraphics[width=8cm]{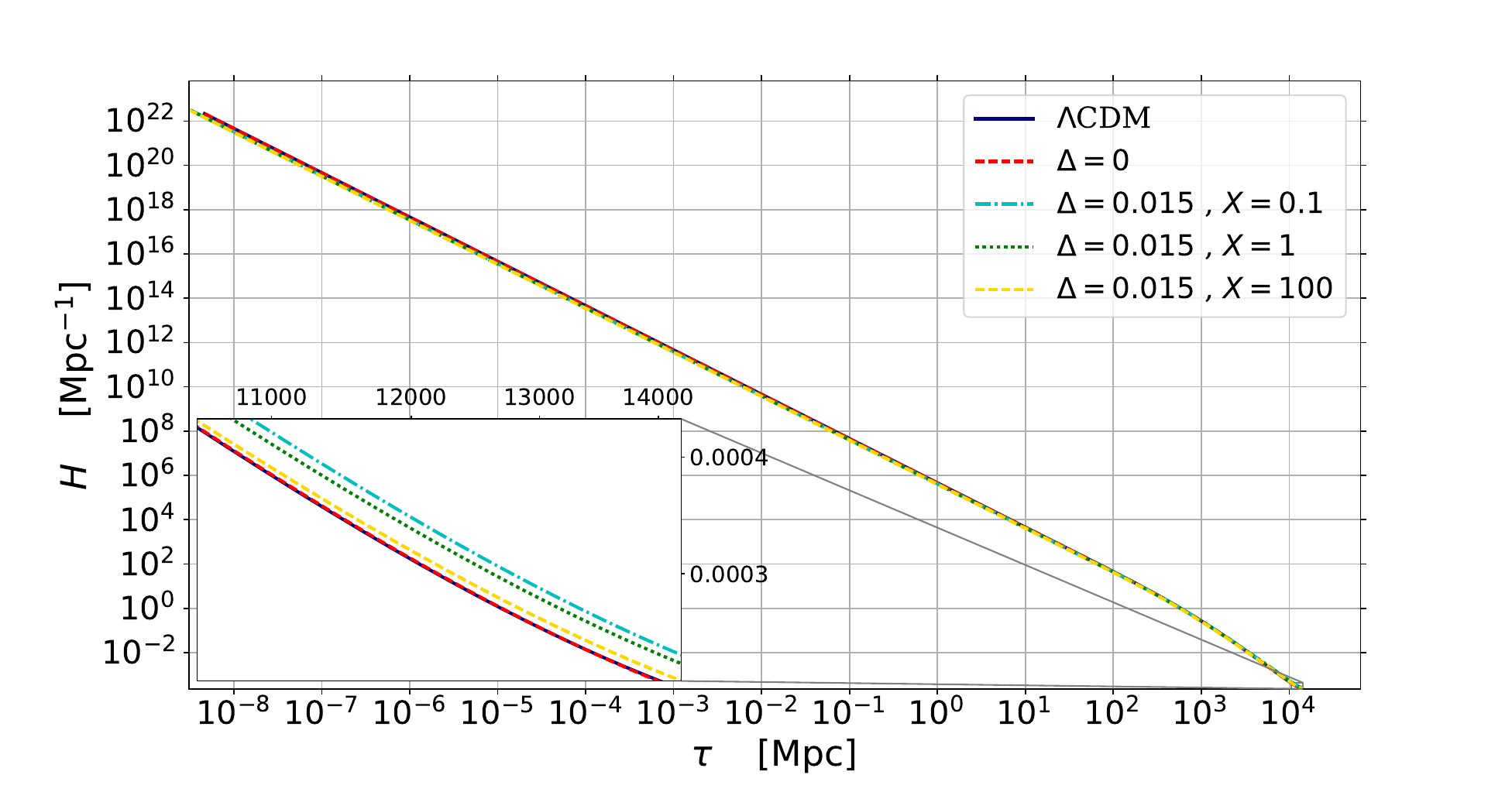}
		\caption{Evolution of Hubble parameter with regard to conformal time in Barrow cosmology compared to $\Lambda$CDM model, for different values of $\Delta$, while $X=1$ (left panel), and analogous diagrams for different values of $X$, considering $\Delta=0.015$ (right panel).}
		\label{fig3}
	\end{figure}

	Regarding the importance of the evolution of Hubble parameter in Barrow cosmology, figure \ref{fig3} displays the expansion history of the universe as described in the modified Friedmann equation (\ref{eq17}). Considering figure \ref{fig3}, it is evident that there is an enhancement in the current value of Hubble parameter for non zero values of $\Delta$, which shows more consistency with low-redshift estimations of Hubble constant \cite{H03,H04,H05,H06,H07}.   
	
	Accordingly, for more precise investigations on the capability of Barrow model in reducing the Hubble tension, we fit the model to current and future observational data by performing an MCMC\footnote{Markov Chain Monte Carlo} method through the M\textsc{onte} P\textsc{ython} code \cite{mp1,mp2}. In order to use the mock LISA SS data in our MCMC analysis, we add three created likelihoods based on Pop III, Delay, and No Delay population models to the M\textsc{onte} P\textsc{ython} code. The cosmological parameter set contemplated in MCMC approach includes \{ $100\,\Omega_{\mathrm{B},0} h^2$, $\Omega_{\mathrm{DM},0} h^2$, $100\,\theta_s$, $\ln (10^{10} A_s)$, $n_s$, $\tau_{\mathrm{reio}}$, $w_\mathrm{DE}$, $\Delta$, $X$ \}, which are the six parameters of standard model of cosmology, and also the dark energy equation of state $w_\mathrm{DE}$, the deformation exponent $\Delta$, and the constant $X$. There are also some derived parameters in MCMC analysis which consist of the reionization redshift $z_\mathrm{reio}$, the matter density parameter $\Omega_{\mathrm{M},0}$, the Hubble constant $H_0$, and the structure growth parameter $\sigma_8$. Moreover, preliminary numerical investigations suggest the prior ranges [$0$, $0.015$] and [$0.1$, $100$] for $\Delta$ and $X$, respectively, while we consider no prior range for $w_\mathrm{DE}$.
	
\begin{figure}[h!]
		\centering
		\includegraphics[width=8.5cm]{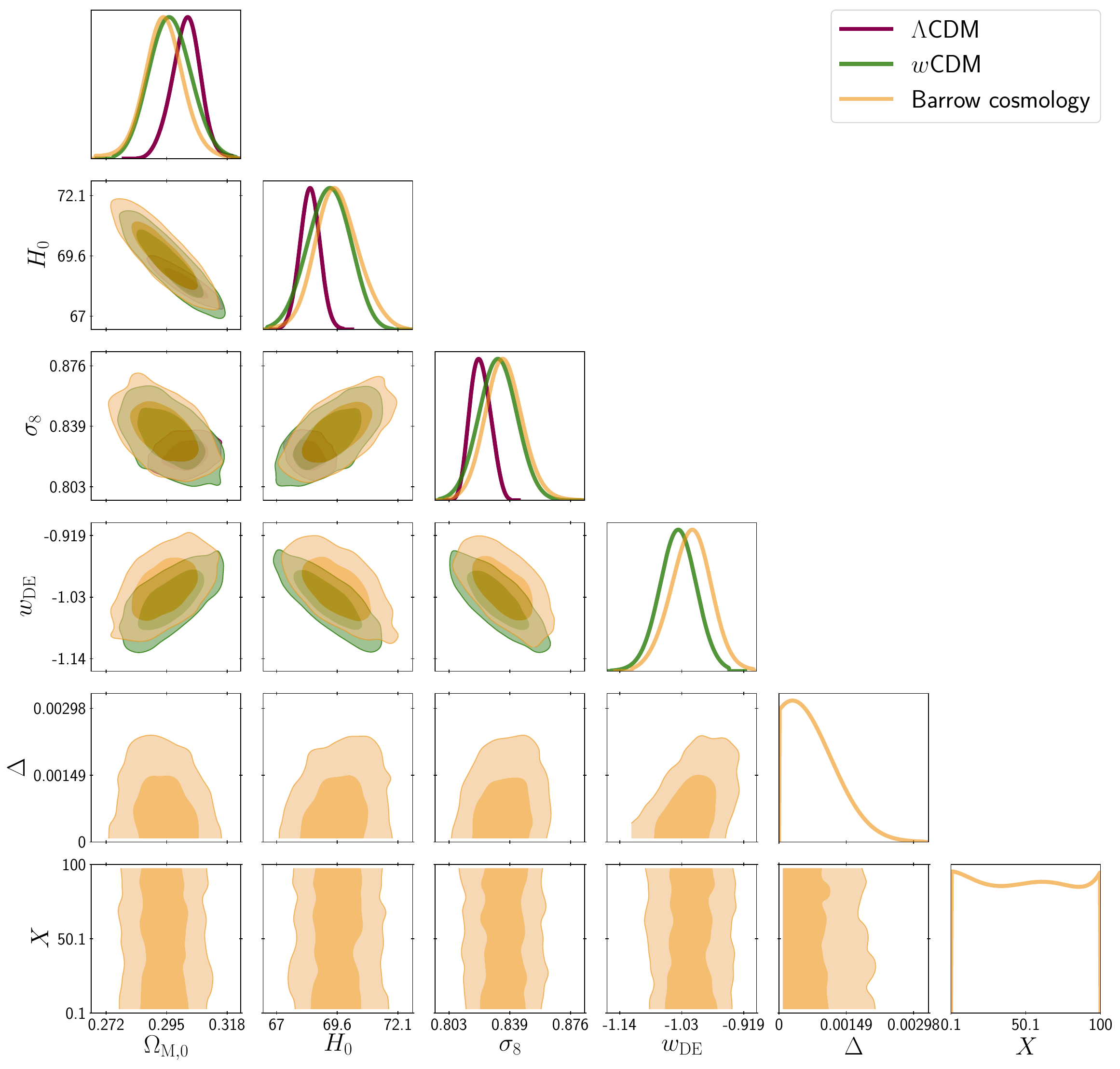}
		\caption{One-dimensional marginalized posterior and two-dimensional marginalized contours with $68\%$ and $95\%$ confidence levels from "CMB+SN+BAO" dataset, for some selected parameters of Barrow model (orange) compared to $\Lambda$CDM (purple) and $w$CDM (green).}
		\label{fig4}
	\end{figure}

	In pursuance of deriving constraints on cosmological parameters of Barrow model, we first apply the "CMB+SN+BAO" dataset, where the results for Barrow cosmology compared to our base models, namely the standard $\Lambda$CDM model and the $w$CDM model (described by cold dark matter and dark energy fluid with constant equations of state) are reported in table \ref{tab1}. 	
	\begin{table}[h!]
		\centering
		\caption{Best fit values along with $1\sigma$ and $2\sigma$ confidence level intervals from the "CMB+SN+BAO" dataset for $\Lambda$CDM, $w$CDM, and Barrow cosmology.}
		\scalebox{.65}{
			\begin{tabular}{|c|c|c|c|c|c|c|} 
				\hline    			 
				& \multicolumn{2}{|c|}{} & \multicolumn{2}{|c|}{} & \multicolumn{2}{|c|}{} \\
				& \multicolumn{2}{|c|}{$\Lambda$CDM} & \multicolumn{2}{|c|}{$w$CDM} & \multicolumn{2}{|c|}{Barrow cosmology} \\
				\cline{2-7}  
				& & & & & & \\
				{parameter} & best fit & 68\% \& 95\% limits & best fit & 68\% \& 95\% limits & best fit & 68\% \& 95\% limits \\ \hline 
				& & & & & & \\
				$100\,\Omega_{\mathrm{B},0} h^2$ & $2.246$ & $2.245^{+0.013+0.027}_{-0.014-0.028}$ & $2.238$ & $2.239^{+0.014+0.028}_{-0.014-0.028}$ & $2.236$ & $2.240^{+0.014+0.030}_{-0.015-0.028}$ \\ 
				& & & & & & \\
				$\Omega_{\mathrm{DM},0} h^2$ & $0.1191$ & $0.1189^{+0.00096+0.0018}_{-0.00086-0.0020}$ & $0.1198$ & $0.1194^{+0.0011+0.0020}_{-0.00099-0.0021}$ & $0.1196$ & $0.1198^{+0.0011+0.0023}_{-0.0011-0.0023}$ \\
				& & & & & & \\
				$100\,\theta_s$ & $1.042$ & $1.042^{+0.00029+0.00057}_{-0.00030-0.00059}$ & $1.042$ & $1.042^{+0.00029+0.00057}_{-0.00031-0.00056}$ & $1.042$ & $1.042^{+0.00029+0.00059}_{-0.00030-0.00060}$ \\
				& & & & & & \\
				$\ln (10^{10} A_s)$ & $3.048$ & $3.047^{+0.013+0.028}_{-0.014-0.028}$ & $3.040$ & $3.045^{+0.014+0.030}_{-0.014-0.027}$ & $3.041$ & $3.047^{+0.013+0.027}_{-0.015-0.027}$ \\
				& & & & & & \\
				$n_s$ & $0.9675$ & $0.9673^{+0.0036+0.0076}_{-0.0038-0.0074}$ & $0.9647$ & $0.9660^{+0.0040+0.0081}_{-0.0041-0.0078}$ & $0.9682$ & $0.9670^{+0.0042+0.0079}_{-0.0039-0.0084}$ \\
				& & & & & & \\
				$\tau_\mathrm{reio}$ & $0.05699$ & $0.05662^{+0.0063+0.014}_{-0.0073-0.013}$ & $0.05223$ & $0.05536^{+0.0061+0.014}_{-0.0076-0.014}$ & $0.05364$ & $0.05519^{+0.0063+0.013}_{-0.0076-0.014}$  \\
				& & & & & & \\
				$w_{\mathrm{DE}}$ & --- & --- & $-1.018$ & $-1.034^{+0.035+0.065}_{-0.032-0.067}$ & $-1.042$ & $-1.012^{+0.036+0.074}_{-0.036-0.071}$ \\
				& & & & & & \\
				$\Delta$ & --- & --- & --- & --- & $0.0002865$ & $0.0007894^{+0.00020+0.0012}_{-0.00079-0.00079}$ \\
				& & & & & & \\
				$X$ [Mpc] & --- & --- & --- & --- & $72.61$ & unconstrained \\
				& & & & & & \\
				$z_\mathrm{reio}$ & $7.921$ & $7.870^{+0.64+1.4}_{-0.72-1.3}$ & $7.463$ & $7.757^{+0.64+1.4}_{-0.73-1.4}$ & $7.606$ & $7.745^{+0.67+1.3}_{-0.73-1.4}$ \\
				& & & & & & \\
				$\Omega_{\mathrm{M},0}$ & $0.3038$ & $0.3022^{+0.0057+0.011}_{-0.0050-0.011}$ & $0.3015$ & $0.2963^{+0.0088+0.016}_{-0.0076-0.016}$ & $0.2910$ & $0.2939^{+0.0078+0.016}_{-0.0081-0.016}$ \\
				& & & & & & \\
				$H_0\;[\mathrm{km\,s^{-1}\,Mpc^{-1}}]$ & $68.28$ & $68.41^{+0.37+0.90}_{-0.46-0.82}$ & $68.67$ & $69.20^{+0.88+1.9}_{-0.97-1.7}$ & $69.84$ & $69.58^{+0.91+1.9}_{-0.94-1.8}$ \\
				& & & & & & \\
				$\sigma_8$ & $0.8225$ & $0.8217^{+0.0056+0.012}_{-0.0061-0.011}$ & $0.8264$ & $0.8319^{+0.011+0.024}_{-0.012-0.022}$ & $0.8372$ & $0.8359^{+0.012+0.022}_{-0.012-0.023}$ \\
				& & & & & & \\
				\hline    
			\end{tabular}
		}    		
		\label{tab1}
	\end{table} 
	Also, the $1\sigma$ and $2\sigma$ confidence contours based on "CMB+SN+BAO" dataset are demonstrated in figure \ref{fig4}. 
	Numerical results exhibit an increase in the value of Hubble constant in Barrow model compared to $\Lambda$CDM, which is positively influenced by the phantom nature of dark energy. Thus, we observe a slight alleviation in Hubble tension for Barrow cosmology with a phantom dark energy, while the tension can not be completely released. On the other hand, the constant $X$ is degenerate with other model parameters, and remains unconstrained for this dataset.
	Moreover, since by choosing $\Delta=0$, Barrow cosmology will reduce to $w$CDM model, it is important to compare Barrow cosmology with the base model $w$CDM as illustrated in table \ref{tab1}. We find that the phantom behaviour of dark energy in $w$CDM model provides a minor refinement in Hubble tension compared to the standard model of cosmology, which represents a reasonable agreement between Barrow cosmological model and $w$CDM.
	
\begin{figure}[h!]
		\centering
		\includegraphics[width=8.5cm]{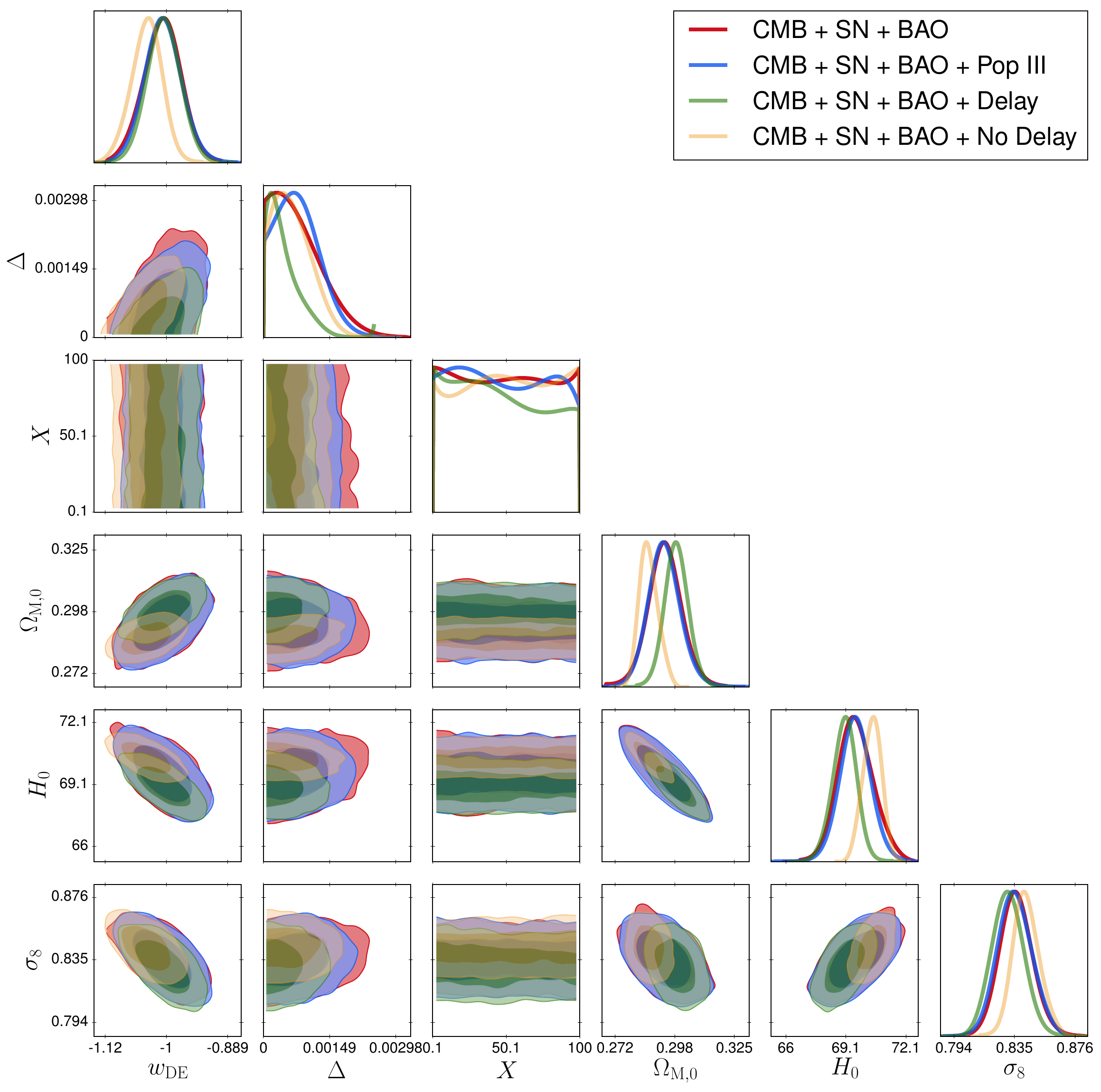}
		\caption{One-dimensional marginalized posterior and two-dimensional marginalized contours with $68\%$ and $95\%$ confidence levels from "CMB+SN+BAO" (red), "CMB+SN+BAO+Pop III" (blue), "CMB+SN+BAO+Delay" (green), and "CMB+SN+BAO+No Delay" (orange) datasets, for some selected parameters of Barrow model.}
		\label{fig5}
	\end{figure}

	Considering the derived best fit values for cosmological parameters of Barrow model in table \ref{tab1}, it is possible to generate three LISA SS mock catalogs, namely, Pop III, Delay, and No Delay, based on the fiducial Barrow cosmology, as displayed in figure \ref{fig1}. Thereupon, we are able to confront Barrow model with SS data, by employing three different datasets combinations: "CMB+SN+BAO+Pop III", "CMB+SN+BAO+Delay", and "CMB+SN+BAO+No Delay". Correspondingly, observational constraints on cosmological parameters of Barrow model are summarized in table \ref{tab2}. 
	\begin{table}[h!]
		\centering
		\caption{Best fit values along with $1\sigma$ and $2\sigma$ confidence level intervals from "CMB+SN+BAO+Pop III", "CMB+SN+BAO+Delay", and "CMB+SN+BAO+No Delay" datasets, regarding Barrow cosmology.}
		\scalebox{.65}{
			\begin{tabular}{|c|c|c|c|c|c|c|} 
				\hline    			 
				& \multicolumn{2}{|c|}{} & \multicolumn{2}{|c|}{} & \multicolumn{2}{|c|}{} \\
				& \multicolumn{2}{|c|}{CMB+SN+BAO+Pop III} & \multicolumn{2}{|c|}{CMB+SN+BAO+Delay} & \multicolumn{2}{|c|}{CMB+SN+BAO+No Delay} \\
				\cline{2-7}  
				& & & & & & \\
				{parameter} & best fit & 68\% \& 95\% limits & best fit & 68\% \& 95\% limits & best fit & 68\% \& 95\% limits \\ \hline 
				& & & & & & \\
				$100\,\Omega_{\mathrm{B},0} h^2$ & $2.238$ & $2.241^{+0.015+0.030}_{-0.015-0.032}$ & $2.230$ & $2.238^{+0.013+0.028}_{-0.014-0.027}$ & $2.247$ & $2.241^{+0.014+0.026}_{-0.014-0.027}$ \\ 
				& & & & & & \\
				$\Omega_{\mathrm{DM},0} h^2$ & $0.1200$ & $0.1197^{+0.0010+0.0023}_{-0.0012-0.0021}$ & $0.1201$ & $0.1198^{+0.0011+0.0023}_{-0.0011-0.0022}$ & $0.1194$ & $0.1197^{+0.00096+0.0021}_{-0.0012-0.0021}$ \\
				& & & & & & \\
				$100\,\theta_s$ & $1.042$ & $1.042^{+0.00030+0.00056}_{-0.00027-0.00062}$ & $1.042$ & $1.042^{+0.00028+0.00057}_{-0.00030-0.00057}$ & $1.042$ & $1.042^{+0.00029+0.00056}_{-0.00028-0.00056}$ \\
				& & & & & & \\
				$\ln (10^{10} A_s)$ & $3.048$ & $3.047^{+0.013+0.028}_{-0.016-0.029}$ & $3.046$ & $3.045^{+0.012+0.029}_{-0.017-0.027}$ & $3.049$ & $3.045^{+0.013+0.027}_{-0.014-0.028}$ \\
				& & & & & & \\
				$n_s$ & $0.9661$ & $0.9673^{+0.0044+0.0079}_{-0.0039-0.0082}$ & $0.9659$ & $0.9664^{+0.0040+0.0083}_{-0.0042-0.0082}$ & $0.9647$ & $0.9673^{+0.0038+0.0080}_{-0.0038-0.0077}$ \\
				& & & & & & \\
				$\tau_\mathrm{reio}$ & $0.05578$ & $0.05575^{+0.0065+0.014}_{-0.0081-0.014}$ & $0.05699$ & $0.05476^{+0.0061+0.014}_{-0.0082-0.014}$ & $0.05648$ & $0.05486^{+0.0062+0.014}_{-0.0075-0.014}$  \\
				& & & & & & \\
				$w_{\mathrm{DE}}$ & $-1.012$ & $-1.010^{+0.036+0.071}_{-0.035-0.072}$ & $-1.032$ & $-1.009^{+0.030+0.065}_{-0.033-0.062}$ & $-1.057$ & $-1.042^{+0.029+0.060}_{-0.030-0.060}$ \\
				& & & & & & \\
				$\Delta$ & $0.0005820$ & $0.0007850^{+0.00032+0.00089}_{-0.00065-0.00078}$ & $0.0002594$ & $0.0004571^{+0.00011+0.00071}_{-0.00046-0.00046}$ & $0.0005151$ & $0.0006632^{+0.00021+0.00079}_{-0.00063-0.00066}$ \\
				& & & & & & \\
				$X$ [Mpc] & $38.51$ & unconstrained & $41.08$ & unconstrained & $1.712$ & unconstrained \\
				& & & & & & \\
				$z_\mathrm{reio}$ & $7.829$ & $7.796^{+0.68+1.4}_{-0.79-1.4}$ & $7.968$ & $7.706^{+0.65+1.4}_{-0.79-1.4}$ & $7.858$ & $7.702^{+0.64+1.4}_{-0.74-1.4}$ \\
				& & & & & & \\
				$\Omega_{\mathrm{M},0}$ & $0.2971$ & $0.2936^{+0.0074+0.015}_{-0.0075-0.015}$ & $0.2972$ & $0.2989^{+0.0056+0.011}_{-0.0057-0.011}$ & $0.2841$ & $0.2867^{+0.0041+0.0084}_{-0.0045-0.0085}$ \\
				& & & & & & \\
				$H_0\;[\mathrm{km\,s^{-1}\,Mpc^{-1}}]$ & $69.23$ & $69.58^{+0.89+1.8}_{-0.86-1.8}$ & $69.21$ & $68.98^{+0.64+1.3}_{-0.63-1.3}$ & $70.68$ & $70.40^{+0.45+0.97}_{-0.52-0.95}$ \\
				& & & & & & \\
				$\sigma_8$ & $0.8355$ & $0.8352^{+0.012+0.024}_{-0.012-0.024}$ & $0.8376$ & $0.8309^{+0.012+0.021}_{-0.011-0.022}$ & $0.8452$ & $0.8428^{+0.0096+0.019}_{-0.0092-0.019}$ \\
				& & & & & & \\
				\hline    
			\end{tabular}
		}    		
		\label{tab2}
	\end{table} 
	Furthermore, two-dimensional marginalized posterior distributions for selected cosmological parameters of Barrow model are represented in figure \ref{fig5}.
	 
	Regarding numerical investigations on Barrow cosmology, we find that adding the mock GW SS data to current cosmological probes can provide better constraints on the model parameters. In particular, we notice that the exponent $\Delta$ is partially better constrained after the addition of Pop III data, while including Delay data would result in tighter constraints on the most of cosmological parameters. 
	On the other hand, when we add No Delay data, much tighter constraints on the model parameters would be obtained. Moreover, the addition of No Delay data is also more effective in ameliorating the Hubble tension by predicting higher values for $H_0$, which is certainly affected by the phantom behavior of the dark energy equation of state. 
	
	It is also interesting to consider the measurement precision of the Hubble constant defined as $\sigma(H_0)/H_0$, where $\sigma(H_0)$ is the $1\sigma$ error of Hubble constant. According to MCMC analysis, the obtained measurement precision of $H_0$ based on "CMB+SN+BAO" dataset is $2.66\%$, where adding the Pop III data would not improve it significantly by providing the constraint precision of $2.52\%$. However, in case of "CMB+SN+BAO+Delay" dataset the constraint precision of $H_0$ would improve to $1.84\%$, while for the "CMB+SN+BAO+No Delay" dataset we measure a considerable precision of $1.38\%$. Thus, it can be concluded that the mock LISA SS data is capable of providing noticeable precision to constrain Barrow cosmological model.    	
	\section{Conclusions} \label{sec5}
	Barrow cosmology is based on Barrow corrections to the area law entropy of the apparent horizon of the universe, resulted from quantum-gravitational deformation effects. Regarding the prominent analogy between thermodynamics and gravity, one can drive modified field equations of Barrow model by associating the Barrow entropy to the apparent horizon of FLRW universe, as described in section \ref{sec2}, where the propagation of GWs in Barrow cosmology is also investigated. The purpose of the present paper is to study the Barrow cosmological model with observational probes, including current data along with the mock LISA SS data. 
	
	LISA is a future space-borne GW observatory, designed to detect GW signals from MBHBs in the milli-Hz frequency band. We generate three SS catalogs, namely Pop III, Delay, and No Delay, according to ten years of LISA mission duration, where the Barrow cosmology constrained by the "CMB+SN+BAO" dataset is contemplated as the fiducial model.
	
	Numerical investigations based on dataset "CMB+SN+BAO" indicate that Barrow model with a phantom dark energy can slightly relieve the Hubble tension, while the problem cannot be solved completely. Furthermore, when we add LISA SS data to current observations, constraints on the model parameters would be improved. Specifically, in case of "CMB+SN+BAO+Pop III" dataset, better constraints on the deformation exponents $\Delta$ is resulted. Moreover, the model parameters are constrained more tightly for the dataset "CMB+SN+BAO+Delay". In addition, we observe significantly tighter constraints on cosmological parameters in case of "CMB+SN+BAO+No Delay" dataset, and also, the Hubble tension is reduced more impressively in this case. 
    \section*{Code availability}
    The modified version of the CLASS code is available under reasonable request.
    \section*{Acknowledgements}
    DFM thanks the Research Council of Norway for their support and the resources provided by 
    UNINETT Sigma2 -- the National Infrastructure for High Performance Computing and 
    Data Storage in Norway.  AA acknowledges SarAmadan and school of physics of IPM.

\bibliographystyle{JHEP}
\bibliography{ref}

\end{document}